# [NE V] EMISSION IN OPTICALLY CLASSIFIED STARBURSTS

N. P. Abel[1] & S. Satyapal[2]

[1]Department of Physics, University of Cincinnati, Cincinnati, OH, 45221
npabel2@gmail.com

[2]Department of Physics and Astronomy, George Mason University, Fairfax, VA 22030-4444, satyapal@physics.gmu.edu

## Abstract

Detecting Active Galactic Nuclei (AGN) in galaxies dominated by powerful nuclear star formation and extinction effects poses a unique challenge. Due to the longer wavelength emission and the ionization potential of $Ne^{3+}$, infrared [Ne V] emission lines are thought to be excellent AGN diagnostics. However, stellar evolution models predict Wolf-Rayet stars in young stellar clusters emit significant numbers of photons capable of creating $Ne^{4+}$. Recent observations of [Ne V] emission in optically classified starburst galaxies require us to investigate whether [Ne V] can arise from star-formation activity and not an AGN. In this work, we calculate the optical and IR spectrum of gas exposed to a young starburst and AGN SED. We find: 1) a range of parameters where [Ne V] emission can be explained solely by star-formation and 2) a range of relative AGN to starburst luminosities that reproduces the [Ne V] observations, yet leaves the optical spectrum looking like a starburst. We also find infrared emission-line diagnostics are much more sensitive to the AGN than optical diagnostics, particularly for weak AGN. We apply our model to the optically classified, yet [Ne V] emitting, starburst galaxy NGC 3621. We find, when taking the infrared and optical spectrum into account, ~30-50% of the galaxy's total luminosity is due to an AGN. Our calculations show that [Ne V] emission is almost always the result of AGN activity. The models presented in this work can be used to determine the AGN contribution to a galaxy's power output.

## Introduction & Background

The importance of Active Galactic Nuclei (AGN) and star-formation (or starbursts) to the overall energy output of a galaxy is an important science question. In many galaxies, both AGN and starburst activity contribute an appreciable fraction of the total galaxy luminosity (Genzel et al. 1998). Discriminating between the two tells us which physical process – accretion onto a Supermassive Black Hole (SBH) or star-formation out of the Interstellar Medium (ISM) – dominates the energetics of the galaxy. In turn, answering this question tells us about the physical process which heats the gas and dust along with producing the emission, absorption, and dust continuum spectrum. Therefore, determining "How much AGN vs. how much Starburst?" allows us to better interpret spectra that has been or will be observed by a variety of telescopes such as Hubble, Spitzer, Herschel, SOFIA, ALMA, and James Webb.

Several ways exist to determine the relative AGN and starburst activity in a galaxy. Most involve the combination of spectroscopic observations combined with a model of the Spectral Energy Distribution (SED) of an AGN. These methods utilize the primary physical difference between an SED of an AGN and starburst. In an AGN, the SED is much harder (more high energy photons) than a stellar SED, and will therefore excite emission-lines from higher ionization states than a starburst.

Such analysis is not always straightforward. The presence of shocks, extinction effects, our incomplete understanding of the AGN SED, and the degeneracy of a weak AGN with compact nuclear starburst makes the job of disentangling AGN from star-formation activity ambiguous. In this work, we investigate the role of AGN and starbursts on the observed emission line spectrum, with particular emphasis on optically classified starbursts with observed infrared AGN properties. Section 2 reviews optical and infrared spectroscopic methods used in discriminating between AGN and starburst, Section 3 describes our computational model, Section 4 presents the results of our calculations, and Section 5 compares our calculations to NGC 3621. We end with a series of conclusions in Section 6.

# 1 How Much AGN, How Much Starburst

Studying the role of the AGN in a galaxy is limited by many factors. In galaxies where star-formation contributes the bulk of the bolometric luminosity, the AGN is relatively weak and hard to detect spectroscopically. Classic optical diagnostic diagrams (Veilleux & Osterbrock 1987) along with strong PAH features in the infrared (Genzel et al. 1998) can both suggest that such galaxies are starburst-like. In the X-ray regime, a weak AGN can be indistinguishable from X-ray binaries in the host galaxy, while radio emission is often times indistinguishable from a compact nuclear starburst. In the following paragraphs, we summarize the optical and infrared methods for detecting AGNs, which are the focus of the rest of this work.

The most common approach to discriminate between AGN and starburst activity is to use optical diagnostic diagrams (Veilleux & Osterbrock 1987). Such plots show two observed emission line ratios on a single plot, such as [O III] $\lambda 5007$Å/H$\beta$, [N II] $\lambda 6583$Å /H$\alpha$, [S II] ($\lambda 6716$Å + $\lambda 6731$Å)/H$\alpha$, and [O I] $\lambda 6300$Å /H$\alpha$. These plots show a clear separation between observations of starburst and AGN, although many galaxies fall in the range where the designation of a galaxy as starburst or AGN is ambiguous. Kewley et al. (2001), using the MAPPINGS III code (Sutherland & Dopita 1993), theoretically calculated the optical spectrum of gas and dust exposed to a starburst continuum and (separately) an AGN. This work then mixed the two calculations together to provide a way to determine the relative AGN/starburst activity.

X-ray and radio observations also provide a useful AGN diagnostic. AGNs emit large amounts of X-rays, which in the non-Compton thick regime provide a powerful tool in detecting optically obscured AGN. However, weak AGN X-ray luminosities will often be indistinguishable from X-ray binaries in the host galaxy (Satyapal et al. 2007). Radio emission from an AGN and a compact nuclear starburst can also be indistinguishable (Condon et al. 1991).

The Mid-IR diagnostic diagrams of Genzel et al. (1998) provide another method to discriminate between AGN and starburst activity, one that is also less sensitive to extinction. The diagnostics of Genzel et al. (1998) rely on two facts. (1) PAH emission is suppressed in AGNs (Roche et al. 1991) and (2) as the level of AGN activity increases, the strength of emission-lines emerging from higher stages of ionization, such as [Ne V] 14.3 µm or [O IV] 25.88µm, increases. Therefore, a plot of increasing PAH emission strength vs. increasing [Ne V] or [O IV] /[Ne II] 12.81µm, AGN and starburst observations will lie in two distinct regions. By making this diagram, and constructing a "mixing-law" that fit the data, Genzel et al. (1998) concluded that about 70-95% of the luminosity coming from UltraLuminous InfraRed Galaxies (ULIRGs) were due to star-formation.

## 1.1 [Ne V]

An infrared spectral feature commonly used to detect the presence of AGN activity is [Ne V] emission, either at 14.3µm or 24.3µm (see, for instance, Groves, Dopita, & Sutherland 2006). The reason [Ne V] emission is thought to arise from an AGN and not star-formation is that the production of $Ne^{4+}$ requires photons with energies greater than ~97eV. Since stars typically do not emit photons beyond 4 Ryd (54.4eV), all [Ne V] emission is attributed to AGN activity. Additionally, [Ne V] is seen in emission towards the center of galaxies, suggesting a link between the physical processes at the center of the galaxy and the formation of [Ne V]. Therefore, [Ne V] provides a powerful diagnostic tool in detecting AGN. Since both [Ne V] lines are in the infrared, the presence of [Ne V] can reveal an AGN even when the AGN is obscured in the optical. This was recently used by Satyapal et al. (2007) to detect an AGN in NGC 3621, a SAd III-IV galaxy which showed no previous evidence for nuclear activity.

Although [Ne V] is an excellent diagnostic for detecting AGN, it is possible that processes other than the hard, non-stellar radiation field produced by an AGN can create $Ne^{4+}$. Shocks are known to play at least some role in the many AGN, even though the combination of observations and detailed calculations suggest photoionization is the dominant ionization process in the Narrow Line Region (NLR) of an AGN (Groves et al. 2004; Ferguson et al. 1997). In particular, Contini et al. (2004) showed the observed infrared emission observed in several AGN can be explained primarily using a shock model. Perhaps more surprisingly, in some circumstances Ne$^{4+}$ (and hence [Ne V]) can be produced by gas exposed to a young stellar population SED (Schaerer & Stasińska 1999). The physical process producing the high energy continuum is the formation of Wolf-Rayet Stars, along with a few massive O stars (Schaerer & Stasińska 1999, C. Leitherer, private communication). This can be clearly seen Figure 1 of Kewley et al. (2001), where the STARBURST99 SED produces significant Extreme UltraViolet (EUV) flux with energies >97 eV. Figure 2 of Schaerer & Stasińska (1999) shows, for a starburst age of 3-4 Myr, the ratio of [Ne V]/[Ne III] 15.6μm exceeds the detection limit for NGC 5253. The same SED seems to explain [O IV] emission, which requires photon energies > 54.9 eV, in some galaxies (Lutz et al. 1998; Schaerer & Stasińska 1999).

While the evidence points to [Ne V] emerging from photoionization by an AGN, recent observations force us to look at the possibility that, under certain conditions, [Ne V] could have a component which is due to star-formation. Satyapal et al. (2007) found, using the Spitzer Space Telescope, towards NGC 3621. This detection is surprising, since NGC 3621 is not optically classified as an AGN (Dale et al. 2006), and because NGC 3621 does not have an observable bulge. This second point is important since galactic bulges are strongly correlated with accreting black holes/AGN activity. Archival Spitzer data (Satyapal et al. 2007, in preparation), show that NGC 3621 is not unique, that there are in fact several galaxies which are (1) optically classified as a normal galaxy (2) contain little or no bulge, yet (3) emit [Ne V]. Is it possible that [Ne V] is not coming from an AGN, but instead a young starburst with a significant W-R and hot O star population?

## 2 Theoretical Calculations

In this section, we outline a series of calculations designed to explore under what conditions observations of [Ne V] emission can be explained by star-formation. Specifically, we calculate the spectrum of gas and dust exposed to a combined AGN/starburst continuum to determine under what conditions a galaxy, optically classified as a starburst, will emit [Ne V]. We will then compare our calculations to the specific case of NGC 3621 to determine if [Ne V] emission is due to AGN activity alone, or if star-formation alone can explain the observations. All our calculations use the spectral synthesis code Cloudy, last described by Ferland et al. (1998). We give only the most important aspects of our computational details, and point the reader to van Hoof et al. (2004), Abel et al. (2005), and references therein for a detailed description of the treatment of important physical processes in Cloudy.

### 2.1 SED

Our calculations are designed to model the spectrum of an H$^+$ region illuminated by radiation emitted by both star-formation and an AGN. We therefore use two different SEDs, one for each physical process, in our calculations. Our choice of star-formation SED is designed to maximize the number of photons emitted above the $Ne^{3+}$ ionization limit, while our AGN continuum is designed to represent what is believed to be a "typical AGN SED." By making this our goal, then we should be able to conclude whether [Ne V] emission can have a stellar component, instead of an AGN.

For the stellar component, we use a 4 Myr continuous star-formation starburst model taken from the Starburst99 website (Leitherer et al. 1999). The parameters of our SED consist of a Saltpeter IMF with a power law of 2.35. All other parameters were left at their default settings from the Starburst99 website. The shape of 4 Myr continuous star-formation SED is similar to an instantaneous star-formation model (Leitherer et al. 1999). Additionally, the WR contribution to the $Ne^{3+}$ ionizing continuum peaks at 4 Myr (Leitherer et al. 1999). When combined with Schaerer & Stasińska (1999) Figure 2, which shows the ratio of [Ne V]/[Ne II] peaking around 4 Myr, our choice of SED is found to achieve its designed intent. We keep the luminosity of the stellar component ($L_{star}$) fixed at $10^{44.6}$ erg s$^{-1}$.

For the AGN component, we use the continuum shape given in Korista et al. (1997), which follows from the work of Elvis et al. (1994). We point the reader to Korista et al. (1997) for a more detailed description of our choice of SED, and highlight the important features here. This continuum consists of an $f_\nu \propto \nu^{-0.5} \exp(-h\nu/kT_{cut})$ UV bump with an X-ray power law proportional to $\nu^{-1}$. For $T_{cut}$, we use the value given in Korista et al. (1997) of $10^{6.0}$ K. The final parameter needed to specify the SED is the ratio of the UV and X-ray continua, $a$(ox), which we fix at $10^{-1.4}$, a typical value for an AGN (Zamorani et al. 1981). We allow $L_{AGN}$ to vary such that the fraction of total $L$ ($L_{total} = L_{star} + L_{AGN}$) due to the AGN ranged from 0.01 to 50%. In addition, we ran a pure AGN model.

## 2.2 Geometry

Our choice of geometry is designed to produce a simple model of gas and dust irradiated simultaneously by two SEDs. We use a 1D plane-parallel geometry, where both the star and AGN are located the same distance away from the illuminated face of the cloud. In an actual galaxy, the AGN would be at the galactic center, while the stellar component would be extended three dimensionally throughout the galaxy. Since this geometry is not computationally feasible, we use a simplified geometry and spend most of our computational power on the microphysical processes which control the emission-line spectrum. By varying both the AGN luminosity and the distance of the SEDs from the illuminated face of the plane-parallel slab (see below), we expect to gain insight into how the AGN can alter the emission-line spectrum, along with its consequences on observations. Additionally, by not having the AGN buried, we fully expose the gas to the AGN, which maximizes the effect of the AGN on the optical emission-line spectrum. A depiction of the assumed geometry is shown in Figure 1.

Other important parameters which define the geometry are the distance of the ionizing SEDs to the illuminated face of the cloud ($R$), the hydrogen density ($n_H$), and the stopping criteria. We chose our value of R so that the dimensionless ratio of ionizing flux to $n_H$, known as the ionization parameter $U$, varied from $10^{0.5}$ to $10^{-4.5}$ in increments of 1 dex. For low LAGN/Ltotal models, this corresponds to $R$ ranging from $10^{20}$ to $10^{22.5}$ cm, in of 0.5 dex increments. Once LAGN becomes significant, the increased contribution of the AGN to the hydrogen-ionizing continuum means we had to increase R needed in order to produce the same dynamical range in $U$. In all models, we use a constant density $n_H = 10^{2.5}$ cm$^{-3}$, a density typical in H$^+$ regions. This choice of $n_H$ is also similar to the density used in Kewley et al. (2001, $n_H$ = 350 cm$^{-3}$). Our combination of SED, $R$, and $n_H$ therefore corresponds to values of $U$ ranging from, which spans the range of $U$ observed in most H$^+$ regions (Veilleux & Osterbrock 1987). In Section 4, we present our results as a function of $U$. We stopped all calculations once the fraction of all hydrogen atoms in the form of H$^+$ falls below 0.1%. This choice of stopping criteria serves two purposes. The stopping criteria allows the calculation to be deep enough to include all emission from the H$^+$, yet shallow enough such that the effects of sphericity are minimal.

## 2.3 Gas and Dust Properties

We assume gas phase abundances consistent with the Orion Nebula, the best studied H$^+$ region. The Orion abundances in Cloudy are defined to be the average abundances taken from Baldwin et al. (1991), Rubin et al. (1991), and Osterbrock, Tran, & Veilleux (1992). For the most important species, the abundances by number are He/H = 0.095, C/H= 3×10$^{-4}$; O/H= 4×10$^{-4}$, N/H= 7×10$^{-5}$, Ne/H= 6×10$^{-5}$, Ar/H= 3×10$^{-6}$, and S/H = 1×10$^{-5}$. For a complete list of abundances, see Ferland (2006). Overall, we include the lightest 30 elements in our calculations, and compute all stages of ionization for each element.

The physics of dust is very important in our models. In addition to being an important heating agent, dust can be an important opacity source of Lyman continuum photons in high $U$ models ($U > 10^{-2}$, Bottorff et al. 1998). We include the effects of dust using a grain size distribution representative of star-forming regions. The ratio of total to selective extinction, $R_V = A_V/(A_B-A_V)$, is a good indicator of the size distribution of grains (Cardelli et al. 1989). Calzetti et al. (2000) derived an average value for $R_V \sim 4.3$. We therefore use the $R_V = 4$ grain size distribution given in Weingartner & Draine (2001). The grain abundance is scaled such that $A_V/N(H_{tot}) = 5\times10^{-22}$ mag cm$^2$. PAHs are thought to be destroyed by hydrogen ionizing radiation and coagulate in molecular environments (see, for instance, Omont 1986). Since we do not go past the H$^+$ region in our calculations, we do not include the PAHs in our model.

With our model assumptions and parameters listed above, we predict the resulting emission-line spectrum as a function of $U$ and $L_{AGN}$. Overall, the combination of $L_{AGN}$ and $R$ represents 30 separate calculations. The results of these calculations are in the subsequent section.

# 3 Discussion

The results of our calculations are shown in Figures 2 and 3, as a function of $U$ and the fraction of the luminosity due to the AGN ($L_{AGN}$ fraction). Figure 2 shows our predictions for the infrared emission-line spectrum; [Ne V] (14.3µm)/$L_B$ (A), [Ne V]/[Ne III] (15.56µm) (B), [Ne V]/[Ne II] (12.81µm) (C), and [O IV] (25.88µm)/[Ne II] (D). Figure 3 shows our predictions for the optical emission-line spectrum; [OIII] (5007Å)/Hβ (A), [OI] (6300Å)/Hα (B), [NII] (6583Å)/Hα (C), [SII] (6716Å + 6731Å) /Hα (D). For the lowest $L_{AGN}$ fraction, all the predicted emission is due to star-formation. Therefore, for constant $U$, changes in the emission-line ratios are entirely due to the increased contribution of $L_{AGN}$ to the total luminosity.

## 3.1 Computational Results

The infrared emission-line ratios show several trends, which are related to changes in the ionization state of the gas brought about by changes in $U$ and $L_{AGN}$. The changes are due to changes in ionization and not temperature because of the weak temperature dependence on IR emission line ratios. In general, Figures 2A, B, and C show an increase in [Ne V] emission relative to $L_B$, [Ne II], and [Ne III] with both $U$ and $L_{AGN}$. Increasing $U$ corresponds to increasing the ratio of photon flux to density, which has the effect of increasing the ionization rate while keeping the recombination rate relatively constant (due to constant density). Increasing $L_{AGN}$ increases the hardness of the radiation field, producing more high-energy photons capable of producing $Ne^{4+}$ ions. For low $L_{AGN}$, the SED is dominated by star-formation. This leads to the $Ne^{4+}/Ne^{2+}$ and $Ne^{4+}/Ne^{+}$ fraction increasing with increasing $U$ over the entire range of $U$ considered. As $L_{AGN}$ increases, the hardness of the radiation field also increases. As a result, increasing $U$ increases $Ne^{4+}/Ne^{2+}$ and $Ne^{4+}/Ne^{+}$ only until $U$ reaches a point where the dominant stage of ionization is $Ne^{5+}$ or higher (see Tarter, Tucker, & Salpeter 1969, Figures 2 & 3). At this point, $Ne^{4+}$ becomes a trace ionization stage, and $Ne^{4+}/Ne^{2+}$ +, $Ne^{4+}/Ne^{+}$ begins to decrease. For [Ne V]/[Ne II], higher $U$ models show a decrease in this ratio when $L_{AGN}$ reaches 5%, while the [Ne V]/[Ne III] ratio decreases when $L_{AGN}$ ~30%. The same argument applies for the [O IV]/[Ne II] ratio, which increases with $U$ for low $L_{AGN}$, but for high $L_{AGN}$ the ratio begins to decrease with increasing $U$ (for high $U$) due to $O^{3+}$ becoming a trace ionization stage.

Figure 3 shows the sensitivity of the optical emission-line ratio diagnostics on $U$ and $L_{AGN}$. The interpretation of Figure 3 is essentially the same as that given in Veilleux & Osterbrock (1987) and we only summarize the essential features here. The [O III]/Hβ ratio (3A) largely increases as $U$ increases, due to the increased ionization of $O^+$. For the largest $U$, however, [O III]/Hβ decreases, which is again due to the changing ionization structure at high $U$ (see above discussion for neon). [OI]/Hα, [NII]/Hα, and [SII]/Hα increase with decreasing $U$, due to the $N^+$, $O^0$, and $S^+$ zones becoming larger as $U$ decreases. As the AGN contribution increases the harder radiation field, particularly in the X-ray regime, creates a larger partially ionized zone with atomic and singly ionized atoms and thereby increasing the emission line ratios.

Figure 2 and 3 shows a range of parameter space where a galaxy can have optical properties characteristic of star-formation, yet also show the presence of [Ne V]. Figure 3 shows that the optical spectrum is completely unaffected by the AGN until the fraction of $L$ due to $L_{AGN}$ > 1%. The point where the AGN affects the optical spectrum depends on $U$, with $U = 10^{0.5}$ having an AGN contribution at $L_{AGN}$ fractions of 1%, while lower $U$ models are typically not affected significantly until $L_{AGN}$ > 10%. The low $U$ result is consistent with the theoretical mixing-line results of Kewley et al. (2001). Figure 2 shows, however, that the AGN affects the [Ne V] ratio at a smaller fraction of the AGN luminosity, with >1 dex increases in [Ne V]/$L_B$, [Ne V]/[Ne III], and [Ne V]/[Ne II] as the fraction of $L$ due to $L_{AGN}$ goes from 0.01 to 0.1%. Therefore, our calculations show the neon spectrum is very sensitive to the AGN and can show the presence of an AGN even when the optical diagnostics classify a galaxy as a starburst, a result observationally shown in Satyapal et al. (2007).

From Figures 2 and 3, our calculations clearly show, at least in theory, that a star-formation continuum with high $U$ could explain [Ne V] emission. Sturm et al. (2002) found an average [Ne V]/[Ne II] ratio of 0.47, with a range of 0.06 to 2.11, based on 13 galaxies optically identified AGN observed with ISO. Additionally, Satyapal et al. (2007) showed, based on observations by Dudik et al. (2007) and Gorjian et al. (2007), that the [Ne V]/$L_B$ ratio is ~$10^{-4}$. Both observables can be explained solely by a 4 Myr continuous star-formation SED with an extreme ionization parameter between $10^{0.5}$ and $10^{-0.5}$. Since each of the infrared ratios in Figure 2 increase with increasing $U$, and since the optical spectrum is unaffected by $L_{AGN}$ for $L_{AGN}/L$ < 1%, our calculations show, at least theoretically, that a young high-$U$ starburst could explain the detection of [Ne V]. While there are other observables that are clearly incompatible with a high $U$ starburst in the Sturm sample, NGC 3621 or other optically classified starbursts galaxies with [Ne V] could be ambiguous.

## 3.2 Application: Case of NGC 3621

Our calculations have wide utility in interpreting [Ne V] emission in a galaxy. In environments where both star-formation and an AGN contribute to the energetics of the galaxy, our results can be used to estimate the fraction of the total luminosity which is due to the AGN and starburst. As with any combination of models with observation, such an estimate would benefit from having as many observables as possible. Additionally, our calculations also show under what conditions [Ne V] emission can arise purely from a star-formation SED.

The primary advantage of our calculations is that they do not use any mixing-law to combine predictions from an AGN and starburst SED separately. Instead, we study the effects of including both continua in a single calculation. The disadvantage is we have used a simplified geometry and neglected shocks. Future models of gas irradiated by both AGN and star-formation activity which includes a more complex geometry and dynamical processes would prove fruitful.

Perhaps the most important application of our models is in the interpretation of [Ne V] emission from regions where, optically, no AGN is detected. As mentioned in Section 2, Satyapal et al. (2007) detected [Ne V] in NGC 3621, a galaxy which, based on the optical spectrum in Dale et al. (2006) and the theoretical modeling of Kewley et al. (2001), could be completely explained by star-formation. Follow-up analysis (Satyapal et al. 2007, in preparation) has found 7 other late-type galaxies like NGC 3621 which, with the exception of [Ne V], show no other spectroscopic signature characteristic of an AGN.

What is the nature of NGC 3621 and other bulgeless galaxies with [Ne V] emission? How does the AGN contribute to the energetics of the galaxy? Or, is there even an AGN at all? To answer this question, we applied our model to the observations of NGC 3621 given in Satyapal et al. (2007) in this work, and apply the calculations presented here to the other 7 galaxies with [Ne V] detections in Satyapal et al. (2007, in preparation). Each of the infrared ratios plotted in Figure 2 A-D were observed with *Spitzer*, while the optical spectra is published in Dale et al. (2006). The infrared observables are: [Ne V]/$L_B$ ~$10^{-4}$, [Ne V]/[Ne II] = 0.06, [Ne V]/[Ne III] = 0.15, and [O IV]/ [Ne II] 0.23. The 1σ errors in each infrared emission ratio are about a factor of 2. The optical emission line ratios are [OIII]/Hβ ~ 1.5 and [NII]/Hα ~ 0.55, with much smaller error bars, as shown in Dale et al. (2006). The observations are shown in Figures 2 and 3 as horizontal dashed lines.

The combination of the models presented here with the observations of NGC 3621 allows us to place important physical constraints on $U$ and the fraction of $L$ due to $L_{AGN}$. Figure 3C shows that the [NII]/Hα can only be explained for $U$ between $10^{-2.5}$ - $10^{-3.5}$, with a best-fit value of approximately $U = 10^{-3}$. This immediately rules out the observed [Ne V] emission being due to a young starburst, which would require $U > 10^{-0.5}$. However, for the range of $U$ just derived, there is also a range of $L_{AGN}/L_{total}$ that reproduces all the observations. Based on Figure 2, $L_{AGN}/L_{total}$ has to be > 5% ([Ne V]/$L_B$), > 10% ([Ne V]/[Ne II]), > 80% ([Ne V]/[Ne III]), and 1-80% (([O IV]/[Ne II]). This means that all of the infrared emission-line diagnostic ratios are consistent with $L_{AGN}/L_{total} \sim 80\%$ . However, we note that all infrared line ratios with the exception of [Ne V]/[Ne III] are well fitted with $L_{AGN}/L_{total}$ 5-80%. We can also set an upper limit to $L_{AGN}/L_{total}$ based on the [OIII]/H$\beta$ ratio. Figure 3A shows, for $L_{AGN}/L_{total} > \sim 50\%$, the value of $U$ required to explain the observed [OIII]/H$\beta$ ratio falls below $10^{-3.5}$, which is inconsistent with [NII]/H$\alpha$. Therefore, taking all observables into account, we find that NGC 3621 is well characterized with $U = 10^{-2.5} - 10^{-3.5}$ and $L_{AGN}/L_{total}$ = 5 – 50%, with a best fit being $U \sim 10^{-3}$, $L_{AGN}/L_{total} \sim$ 30-50%.

Our results show the power of combining the infrared and optical spectrum, along with the utility of our calculations. Although Dale et al. (2006) placed NGC 3621 in the starburst portion of optical diagram; our calculations show that the energetics of NGC 3621 appears to be equally driven by AGN activity. The detection of other galaxies which appear starburst like (based on their optical spectrum), yet show [Ne V] in emission, have the potential to provide important constraints on the relative AGN-to-starburst contribution in a galaxy. In addition, these types of galaxies can guide future theoretical models, allowing us to better understand how to classify galaxies based on the observed emission-line spectrum. The models presented here focused on the sensitivity of the optical and infrared spectral diagnostics to changes in $U$ and $L_{AGN}/L_{total}$, and were focused on global trends rather than a detailed model of a single galaxy. While our calculations do reasonably well at explaining the contribution of the AGN to NGC 3621, the lack of agreement of [Ne V]/[Ne III] with the rest of the observables shows there is clearly more work to be done. On the observational end, we need more infrared observations of optically classified starbursts with weak AGN to constrain the physical properties of these galaxies and improve statistics. From a theoretical viewpoint, expanding the range of parameter space to include geometrical and dynamical effects along with variations in $U$ and $L_{AGN}/L_{total}$ would prove fruitful.

## 4 Conclusions

In this work, we have undertaken a series of photoionization calculations designed to (1) determine the fractional contribution of an AGN to the total luminosity of a galaxy and (2) determine if it is possible for the faint detection of [Ne V] emission in galaxies which are optically classified as starbursts to be explained by the hard ionizing continuum produced by WR stars in young starbursts. As a result of combining our calculations with observations, we have found:

- For a young starburst with a high ionization parameter ($U > 10^{-0.5}$), the observed [Ne V], [Ne III], [Ne II], and [O IV] emission line ratios can be explained solely by a starburst SED. Since typical ionization parameters are $\sim 10^{-2}$ (Veilleux & Osterbrock 1987; see also Osterbrock & Ferland 2005), we find that [Ne V] remains a good tracer of AGN activity in all but the most extreme environments.

- The optical spectrum is left unaffected by the presence of an AGN as long as the AGN luminosity is less than 1 percent of the total luminosity of the galaxy. For $U < 10^{0.5}$, the optical spectrum is insensitive until the AGN luminosity reaches ~10%.

- The infrared spectrum of [Ne V] and [O IV] is much more sensitive to the AGN, with > 1 dex increases in emission-line ratios as the AGN contribution to the total luminosity goes from 0.01 to 0.1%.

- We use our calculations to find that the observed IR and optical spectrum of the bulgeless galaxy NGC 3621, which is optically classified as a starburst, can be characterized by $U \sim 10^{-3}$ where the AGN and starburst contribute slightly less than half of the total galaxy energetics.

Overall, we find our calculations to be extremely useful in disentangling the role of AGN and starburst in galaxies where [Ne V] is detected, yet the optical spectrum is characteristic of a starburst, such as in many late-type galaxies. As more detections of [Ne V] in late-type galaxies become available (see Satyapal et al. 2007, in preparation), the combination of increased statistics and more detailed modeling will allow us to better understand the role of AGN in optically classified starburst galaxies.

Acknowledgements: This material is based upon work supported by the National Science Foundation under Grant No. 0094050, 0607497 to The University of Cincinnati. NPA also acknowledges the University of Cincinnati for computational support, in addition to the University of Kentucky and Miami of Ohio University for a generous allotment of time on their respective supercomputing clusters. S. Satyapal acknowledges financial support from NASA grant NAG5-11432. We thank the referee for carefully reading the manuscript and for many constructive comments.

# 5 References

Abel, N. P., Ferland, G. J., Shaw, G., & van Hoof, P. A. M. 2005, ApJS, 161, 65

Baldwin, J. A., Ferland, G. J., Martin, P. G., Corbin, M. R., Cota, S. A., Peterson, S. A., Bradley, & M., Slettebak, A. 1991, ApJ, 374, 580

Bottorff, M., Lamothe, J., Momjian, E., Verner, E., Vinković, D., & Ferland, G. J. 1998, PASP, 110, 1040

Calzetti, D., Armus, L., Bohlin, R. C., Kinney, A. L., Koornneef, J., & Storchi-Bergmann, T. 2000, ApJ, 533, 682

Cardelli, J. A., Clayton, G. C., & Mathis, J. S. 1989, ApJ, 345, 245

Condon, J. J., Huang, Z.-P., Yin, Q. F., & Thuan, T. X. 1991, ApJ, 378, 65

Contini, M., Viegas, S. M., Prieto, M. A. 2004, MNRAS, 348, 1065

Dale, D. A. et al. 2006, ApJ, 646, 161

Elvis, M., et al. 1994, ApJS, 95, 1

Ferguson, J. W., Korista, K. T., Baldwin, J. A., & Ferland, G. J. 1997, ApJ, 487, 122

Ferland, G. J., Korista, K. T., Verner, D. A., Ferguson, J. W., Kingdon, J. B., & Verner, E. M. 1998, PASP, 110, 76

Ferland, G. J. 2006, Hazy, a Brief Introduction to Cloudy (Univ. Kentucky Dept. Phys. Astron. Int. Rep.)

Genzel, R. et al. 1998, ApJ, 498, 579

Groves, B., Dopita, M., & Sutherland, R. 2006, A&A, 458, 405

Kewley, L. J., Dopita, M. A., Sutherland, R. S., Heisler, C. A., Trevena, J. 2001, ApJ, 556, 121

Kewley, L. J., Heisler, C. A., Dopita, M. A., & Lumsden, S. 2001, ApJS, 132, 37

Korista, K., Baldwin, J., Ferland, G. J., Verner, D. 1997, ApJS, 108, 401

Leitherer, C. et al. 1999, ApJS, 123, 3

Lutz, D., Kunze, D., Spoon, H. W. W., Thornley, M. D. 1998, A&A, 333, 75

Omont, A. 1986, 164, 159

Osterbrock, D. E., Tran, H. D., Veilleux, S. 1992, ApJ, 389, 305

Roche, P. F., Aitken, D. K., Smith, C. H., & Ward, M. J. 1991, MNRAS, 248, 606

Rubin, R. H., Simpson, J. P., Haas, M. R., & Erickson, E. F. 1991, ApJ, 374, 564

Satyapal, S., Vega, D., Heckman, T., O'Halloran, B., Dudik, R. 2007, ApJ, 63L, 9

Schaerer, D., Stasińska, G. 1999, A&A, 345L, 17

Sutherland, R. S. & Dopita, M. 1993, 1993, ApJS, 88, 253

Tarter, C. B., Tucker, W. H., & Salpeter, E. E. 1969, ApJ, 156, 943

van Hoof, P. A. M., Weingartner, J. C., Martin, P. G., Volk, K., & Ferland, G. J., 2004, MNRAS, 350, 1330

Veilleux, S. & Osterbrock, D. E. 1987, ApJS, 63, 295

Weingartner, J. C. & Draine, B. T. 2001, ApJS, 134, 263

Zamorani, G. et al. 1981, ApJ, 245, 357

# 6 Figures

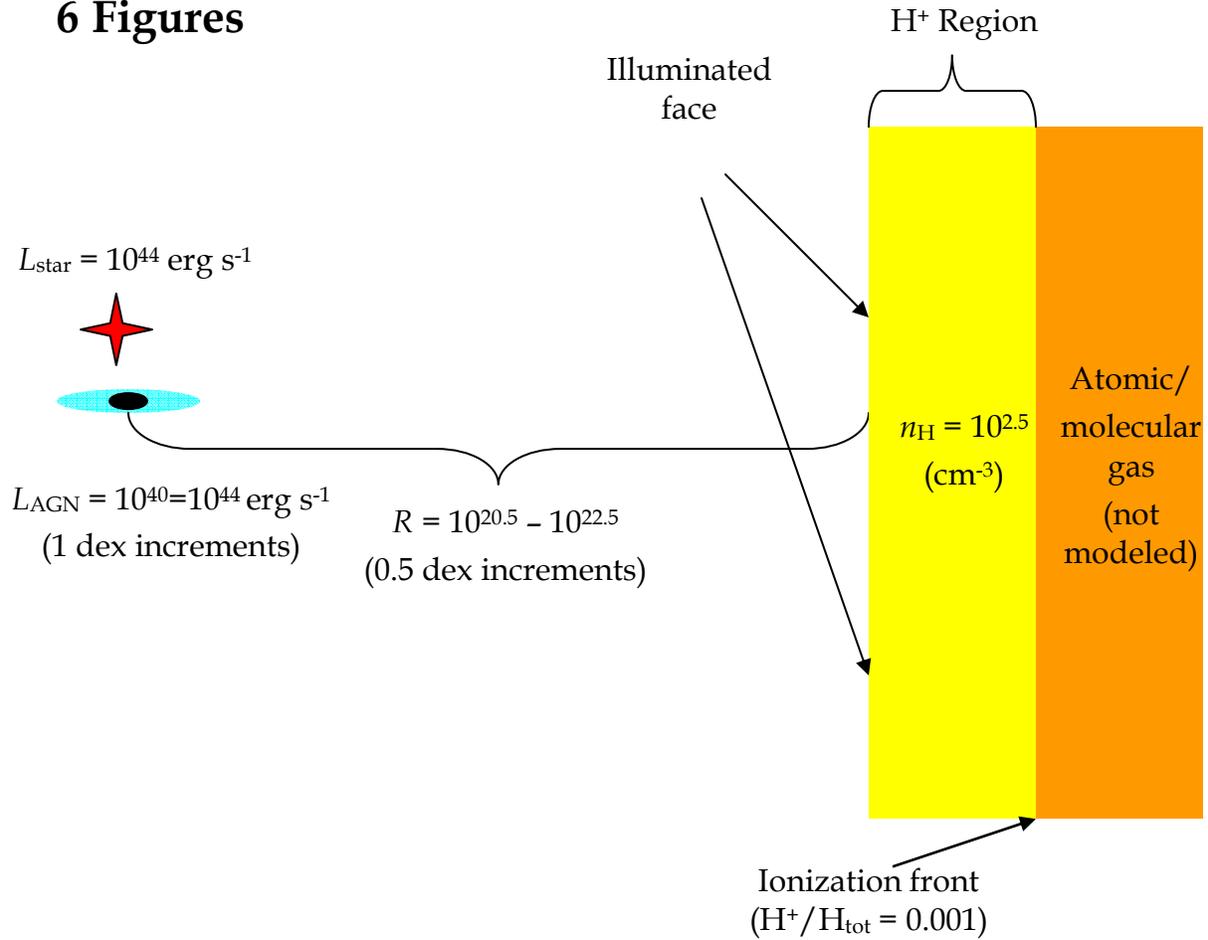

Figure 1 – Geometry used in our calculations. Both an AGN and star-formation SED with luminosity $L$ are located a distance $R$ away from the illuminated face of a plane-parallel slab of gas. We then calculate the emission line spectrum of the $H^+$ region, integrating until we reach the ionization front.

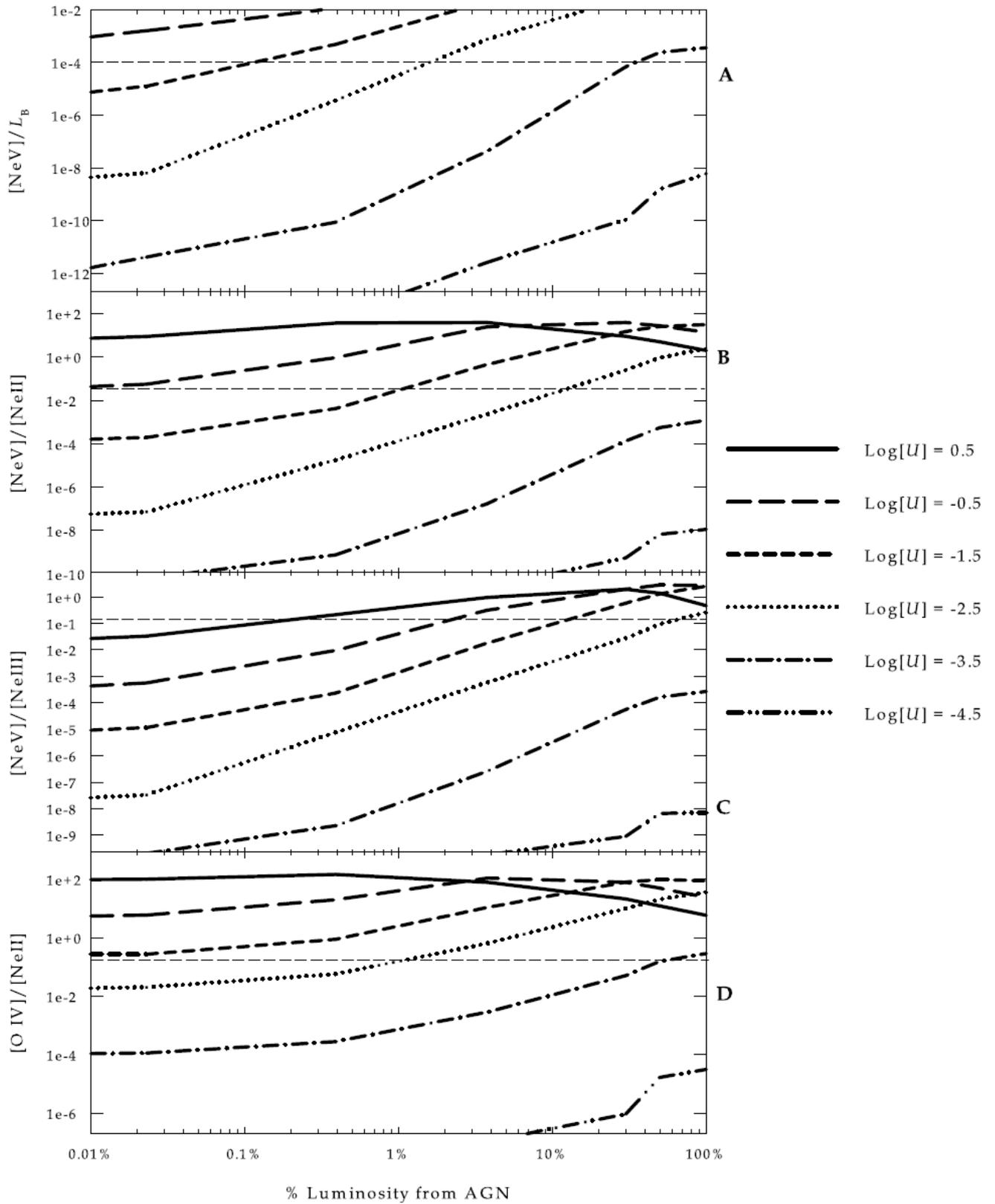

Figure 2 – Predicted infrared emission line ratios, as a function of $U$ and % of luminosity due to an AGN. The horizontal line represents the observations of NGC 3621 by Dale et al. (2006).

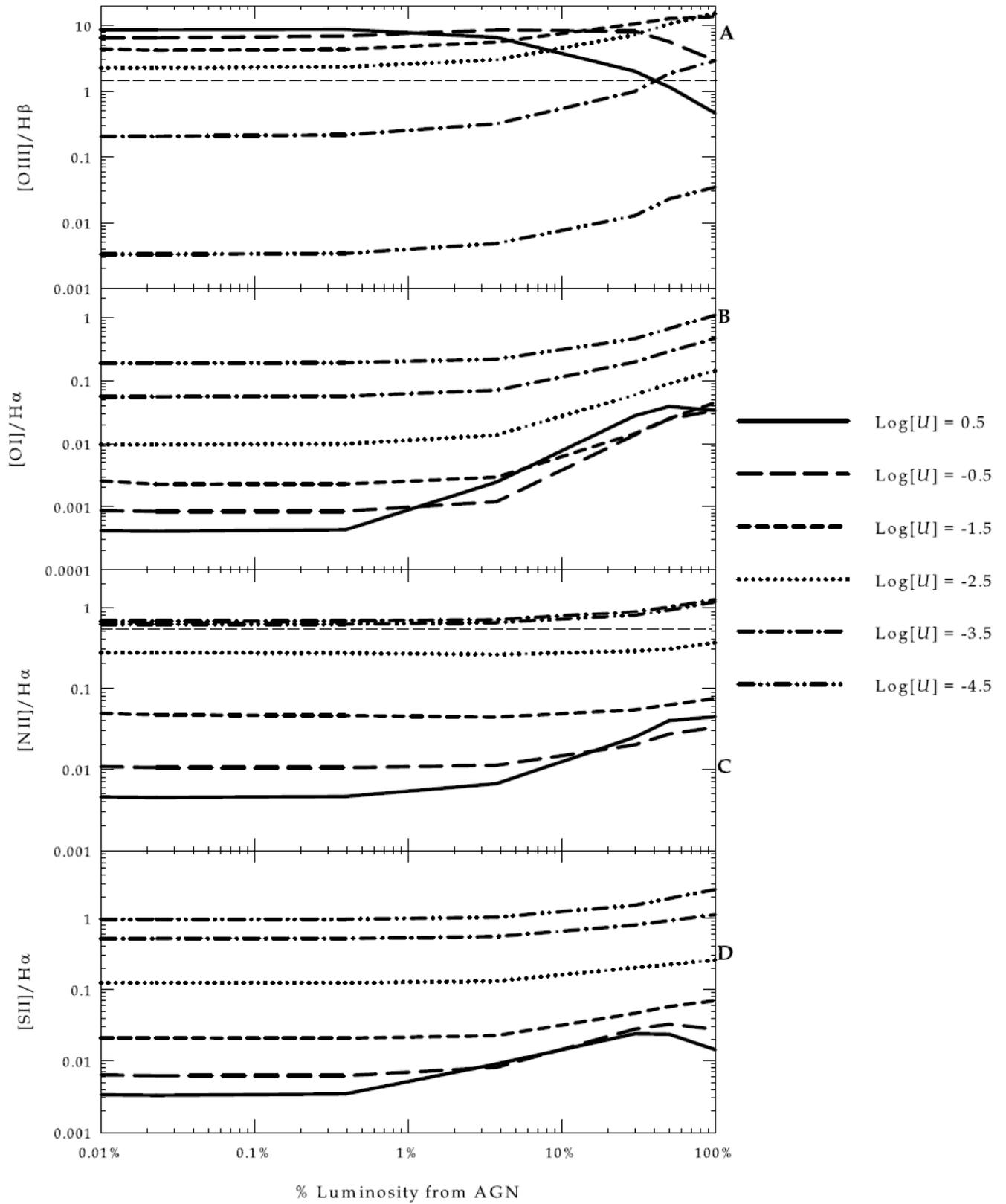

Figure 3– Predicted optical emission line ratios, as a function of $U$ and % of luminosity due to an AGN. The horizontal line represents the observations of NGC 3621 taken from Dale et al. (2006).